\documentclass[12pt]{iopart}

\usepackage{graphicx}
\usepackage{amssymb}
\usepackage{color}
\usepackage{psfrag}
\usepackage{epsfig}
\usepackage{bbm}
\usepackage{bm}
\usepackage{hyperref}
\usepackage{bbold}
\usepackage[normalem]{ulem}
\hypersetup{citecolor=black}
\hypersetup{colorlinks=true}
\hypersetup{linkcolor=blue}
\hypersetup{urlcolor=blue}
\usepackage[latin1]{inputenc} 
\usepackage{soul}
\usepackage[sort&compress,square,comma,numbers]{natbib}

\newcommand{\be}{\begin{equation}}
\newcommand{\ee}{\end{equation}}
\newcommand{\eea}{\end{eqnarray}}

\newcommand{\va}[1]{\ensuremath{(\Delta#1)^2}}

\newcommand{\exs}[1]{\ensuremath{\langle{#1}\rangle}}

\newcommand{\qed}{\ensuremath{\hfill \Box}}

\newcommand{\ket}[1]{\ensuremath{|#1\rangle}}

\newcommand{\kommentar}[1]{}

\newcommand{\aver}[1]{\langle #1 \rangle}

\newcommand{\id}{\openone}

\newcommand{\eqref}[1]{(\ref{#1})}

\newcommand{\SEC}[1]{section~\ref{#1}}
\newcommand{\FIG}[1]{figure~\ref{#1}}

\newcommand{\REF}[1]{\cite{#1}}

\newcommand{\EQ}[1]{\eqref{#1}}

\newcommand{\openone}{\mathbbm{1}}
\newcommand{\tfrac}[2]{\frac{#1}{#2}}
\newcommand{\affiliation}[1]{\address{#1}}
\newcommand{\text}[1]{\mbox{#1}}

\newcommand{\komment}[1]{}

\begin{document}

\title{Entanglement and extreme spin squeezing of unpolarized states}

\author{Giuseppe Vitagliano$^1$, Iagoba Apellaniz$^1$, Matthias Kleinmann$^1$, 
Bernd L\"ucke$^2$, Carsten Klempt$^2$, G\'eza T\'oth$^{1,3,4}$}
\affiliation{$^1$  Department of Theoretical Physics, University of the Basque Country
UPV/EHU, P.O. Box 644, E-48080 Bilbao, Spain}
\affiliation{$^2$ Institut f\"ur Quantenoptik, Leibniz Universit\"at Hannover, Welfengarten 1, D-30167 Hannover, Germany}
\affiliation{$^3$ IKERBASQUE, Basque Foundation for Science, E-48013 Bilbao, Spain}
\affiliation{$^4$ Wigner Research Centre for Physics, Hungarian Academy of Sciences,
P.O. Box 49, H-1525 Budapest, Hungary}
\eads{\mailto{gius.vitagliano@gmail.com, toth@alumni.nd.edu}}

\date{\today}
\begin{abstract}
We present criteria to detect the depth of entanglement in macroscopic  ensembles of spin-$j$ particles using the variance and second moments of the collective spin components.  The class of states detected goes beyond traditional spin-squeezed states by including Dicke states  and other unpolarized states. The criteria derived are easy to evaluate numerically even for systems of very many particles and outperform past approaches,  especially in practical situations where noise is present.  We also derive analytic lower bounds based on the linearization of our criteria, which make it possible to define spin-squeezing parameters for Dicke states. In addition, we obtain spin squeezing parameters also from the condition derived in [A.~S.~S\o rensen and K.~M\o lmer, \href{http://link.aps.org/doi/10.1103/PhysRevLett.86.4431}{Phys. Rev. Lett. 86, 4431 (2001)}]. We also extend our results to systems with fluctuating number of particles.
\end{abstract}




\maketitle


\section{Introduction}

With an interest towards fundamental questions in quantum physics, as
well as applications, larger and larger entangled quantum systems have
been realized with photons, trapped ions, and cold atoms
\cite{Schwemmer2014Experimental, Gao2010Experimental,Leibfried2004Toward,Huang2011Multi-partite,Appel2009Mesoscopic,Sewell2012Magnetic,Gross2010Nonlinear,Lucke2014Detecting,Hosten2016Measurement,Mcconnell2015Entanglement,Haas2014Entangled}. Entanglement is needed for certain quantum information processing
tasks \cite{Horodecki2009Quantum,Guhne2009Entanglement}, and it is also necessary to reach the maximum sensitivity in a
wide range of  interferometric schemes in quantum metrology
\cite{Pezze2009Entanglement}. Hence, the verification of the presence of entanglement is a
crucial  but exceedingly challenging task, especially in an ensemble
of many, say $10^3$ or $10^{12}$ particles \cite{Gross2010Nonlinear,Lucke2014Detecting,Appel2009Mesoscopic,Sewell2012Magnetic,Hosten2016Measurement,Mcconnell2015Entanglement,Haas2014Entangled}. 
Moreover, in such experiments it is 
not sufficient to claim that ``the state is entangled'',
we need also to know how entangled the system is.
Hence, quantifying entanglement in large ensembles has recently been at the
center of attention. In several experiments the {\it entanglement depth}  (i.e., the minimal number of mutually entangled particles consistent with the measurement data)  was determined, reaching to the thousands \cite{Gross2010Nonlinear,Lucke2014Detecting,Hosten2016Measurement,Mcconnell2015Entanglement,Haas2014Entangled}.  

In the many-particle case, especially in large ensembles of cold atoms, it is typically very difficult or 
even impossible to address the particles individually, while measuring 
collective quantities is still feasible.
In this context, one of the most successful approaches to detect entanglement is based on the
criterion \cite{Sorensen2001Many-particle}
\begin{equation}\label{eq:orispinsq}
\xi_{\rm s}^2:= N \frac{(\Delta J_x)^2}{\aver{J_y}^2 +\aver{J_z}^2}\ge 1,
\end{equation}
where $N$ is the number of the spin-$1/2$ particles, $J_l=\sum_{n=1}^N j_l^{(n)}$ for $l=x,y,z$ are the collective spin components,
and $j_l^{(n)}$ are single particle spin components acting on the $n$th particle. Every multiqubit state that violates \EQ{eq:orispinsq} must be entangled \cite{Sorensen2001Many-particle}.
The criterion~\eqref{eq:orispinsq} is best suited for states with a large collective spin in the $(\hat y, \hat z)$-plane and a small variance
$(\Delta J_x)^2$ in the orthogonal direction.
For such states the variance of a spin component is reduced below what can be achieved with fully polarized
spin-coherent  states,  hence they have been called {\it spin squeezed} in the context of metrology \cite{Kitagawa1993Squeezed,Wineland1994Squeezed}. 

As a generalization of \EQ{eq:orispinsq}, a criterion has also been derived by S\o rensen and M\o lmer \cite{Sorensen2001Entanglement}
to detect the depth of entanglement of spin-squeezed states 
in an ensemble of  particles with a spin $j.$
For the criterion, we have to consider a subgroup of $k\le N$ particles and  define its total spin as
\be\label{eq:Jkj}
J=kj.
\ee
We also need to define a function $F_{J}$ via a minimization over quantum states of  such a group as
\begin{equation}\label{fsormolthj}
F_{J}(X):=\frac 1 {J} \min_{\varrho:\frac 1 J \aver{L_z}_\varrho=X} (\Delta L_x)^2_\varrho,
\end{equation}
where $L_l$ are the spin components of the group.
In practice, the minimum will be the same if we carry out the minimization over  states of a single particle with a spin $J$ \cite{Hyllus2012Entanglement}.
Then, for all pure states with an entanglement depth of at most $k$ 
\begin{equation}\label{sormolcrit}
(\Delta J_x)^2 \geq Nj F_{{J}}\left( \frac{\aver{J_z}}{Nj} \right)
\end{equation}
holds. It is easy to see that \EQ{sormolcrit} is valid even for mixed states with an entanglement depth of at most $k$ since the variance is concave in the state and $F_{J}(X)$ is convex
\footnote{The convexity of $F_{J}(X)$ is observed numerically \cite{Sorensen2001Entanglement}. In case the right-hand side of  \EQ{fsormolthj} results in a non-convex function in $X,$ then the convex hull of  the right-hand side of  \EQ{fsormolthj} must be used in the place of $F_J(X)$.}.
Thus, every state that violates \EQ{sormolcrit} must have a depth of entanglement of $(k+1)$ or larger.
It is important to stress that the criterion \EQ{sormolcrit} 
provides a tight lower bound
on $(\Delta J_x)^2$ based on $\aver{J_z}.$
Spin squeezing 
has been demonstrated in many experiments, 
 from cold atoms \cite{Hald1999Spin,Fernholz2008Spin,Orzel2386,Riedel2010Atom-chip-based,Esteve2008Squeezing,Gross2010Nonlinear,BohnetJ2014,Cox2016}  to trapped ions \cite{Meyer2001}, magnetic systems \cite{Auccaise2015} and photons \cite{MitchellExtremespinsqueezing2014},
 and in many of these experiments even multipartite entanglement has been detected using  the condition \eqref{sormolcrit} \cite{Riedel2010Atom-chip-based,Esteve2008Squeezing,Gross2010Nonlinear,BohnetJ2014,Cox2016,MitchellExtremespinsqueezing2014}.
 
Recently, the concept of spin squeezing has been extended to unpolarized states \cite{Toth2007Optimal,He2011Planar,Reis2012,Behbood2014Generation,Vitagliano2014Spin}. In particular,  Dicke states are attracting increasing attention, since their multipartite entanglement is robust against particle loss,
and they can be used for high precision quantum metrology \cite{Lucke2014Detecting}.
Dicke states 
are produced in experiments with photons \cite{Wieczorek2009Experimental,Prevedel2009Experimental} and Bose-Einstein condensates \cite{Lucke2011Twin,Hamley2012Spin-nematic,Lucke2014Detecting}. 
Suitable
criteria to detect the depth of entanglement of Dicke states have also been derived. 
However, either they are limited to spin-$1/2$ particles \cite{Duan2011Entanglement,Lucke2014Detecting} or 
they do not give a tight lower bound on $(\Delta J_x)^2$ based on the  expectation value  measured for the criterion, concretely, $\exs{J_y^2+J_z^2}$ \cite{Zhang2013Generation}.

In this paper, we present a general condition that: (i) provides a lower bound on the entanglement depth, (ii) is applicable to spin-$j$ systems, for any $j$, (iii) works both for spin-squeezed states and Dicke states, 
and, (iv) is close to provide a tight bound in the sense mentioned above in the large particle number limit.
Such a criterion can be applied immediately, for instance, in 
experiments producing Dicke states in spinor condensates \cite{Hoang2016Characterizing}.

We now summarize the main results of our paper.
With a method similar to the one used for obtaining Eq.~(\ref{sormolcrit}),
we show that the condition
\begin{equation}\label{eq:oursormolcritintro}
(\Delta J_x)^2 \geq  Nj  G_J\left(\frac{\aver{J_y^2+J_z^2} - Nj(kj+1)}{N (N-k)j^2}\right) 
\end{equation}
holds for states with an entanglement depth of at most $k$ of an ensemble of $N$ spin-$j$ particles,
where we introduced the notation
\be \label{GJ} G_J : X \mapsto F_J(\sqrt{X}),\ee
with $F_J(X)$ defined as in Eq.~(\ref{fsormolthj}) and $J=kj$ as in \EQ{eq:Jkj}. 
Our approach is motivated by the fact that Eq.~(\ref{sormolcrit}) fails to be a good criterion for mixed states with a low polarization $\aver{J_y}^2 +\aver{J_z}^2\ll N^2j^2$. Thus, we consider the second moments $\aver{J_y^2+J_z^2}$ instead, which are still large for many useful unpolarized quantum states, such as Dicke states. Using the second moments is  advantageous  even for states with a large spin polarization since criteria with second moments are more robust to noise, which will be demonstrated later on concrete examples. We also analyze the performance of our condition
compared to other criteria in the literature.

In general, the function $G_{J}(X)$ appearing on the right-hand side of \EQ{eq:oursormolcritintro} has to be evaluated numerically. However, due to its convexity properties we can bound it from below with 
the  two lowest nontrivial orders of its Taylor expansion around $X=0,$ yielding a spin-squeezing parameter  similar to the one defined in \EQ{eq:orispinsq}. While states saturating \EQ{eq:oursormolcritintro} determine a curve in the $(\aver{J_y^2+J_z^2},(\Delta J_x)^2)$-plane, such an analytic condition corresponds  to tangents to this curve. Hence, we will refer to it  as a linear criterion.
Such a criterion for  states with an entanglement depth $k$ or smaller  is given by the inequality
\begin{equation}\label{eq:ourparameter}
\xi^2:= (kj+1) \frac{2(N-k)j (\Delta J_x)^2}{\aver{J_y^2+J_z^2}-Nj(kj+1)} \geq  1,
\end{equation}
where we require that $kj$ is an integer.
A similar  condition can be obtained from the S\o rensen--M\o lmer criterion (\ref{sormolcrit}) as
\begin{equation}\label{eq:sormolparameter}
\xi^2_{\rm SM}:= (kj+1) \frac{2Nj(\Delta J_x)^2}{\aver{J_y}^2+\aver{J_z}^2} \geq  1,
\end{equation}
again requiring that $kj$ is integer.
A direct comparison between $\xi^2$ and $\xi^2_{\rm SM}$ shows that the former is more suitable for detecting the depth of entanglement of unpolarized states, such as Dicke states.
Note also the similarity between \EQ{eq:sormolparameter} and the original criterion for spin-$1/2$ particles
\EQ{eq:orispinsq}. All these criteria are also generalized to the case when the particle number is not fixed, following~\REF{Hyllus2012Entanglement}.

Our paper is organized as follows. In section 2, we discuss how to evaluate our criteria numerically, while 
we also discuss cases where analytical formulas can be used instead of numerics.  In section 3, we derive our nonlinear entanglement criterion.  In section 4, we present linear criteria leading to new spin-squeezing parameters. In section 5, we compare our entanglement conditions to other conditions existing in the literature. Finally, in section 6, we discuss how to generalize our methods to the case of a fluctuating number of particles.

\section{Numerical computation of $G_J(X)$}
\label{sec:numcomp}

Before describing how to obtain $F_J(X)$ and $G_J(X)$ numerically, we define 
some notions necessary for our discussion.
We distinguish various levels of multipartite entanglement
based on the following definitions.
A pure quantum state is  $k$-producible  if it can be written as 
\be \ket{\psi^{(1)}} \otimes \ket{\psi^{(2)}} \otimes \cdot\cdot\cdot  \otimes \ket{\psi^{(M)}},\label{eq:kprod}\ee
where  $\ket{\psi^{(l)}}$ are states of $k_l \leq k$ particles, and
$M$ stands for the 
number of particle groups. 
A mixed quantum state is  $k$-producible, if it can be written
as a mixture of pure   $k$-producible states.
Clearly, $1$-producible states are separable states.
A state that is not $k$-producible is called $(k+1)$-entangled. 
The entanglement depth is $k+1$ whenever the state is $(k+1)$-producible but not $k$-producible \cite{Guhne2005Multipartite,Sorensen2001Entanglement}. 

Next, we will show a simple method to calculate $F_J(X)$ and $G_J(X).$ We will discuss both numerical and analytical approaches.
Knowing the properties of these functions is necessary to prove later the relation \eqref{eq:oursormolcritintro}.
For an integer $J,$ the function $F_{J}(X)$ given in \EQ{fsormolthj} can be efficiently computed for some interval of $X$ as follows \cite{Sorensen2001Entanglement}. 
We just need to calculate the ground states $|\phi_\lambda\rangle$ of the Hamiltonian
\begin{equation}\label{eq:ssham}
H_\lambda=L_x^2-\lambda L_z 
\end{equation}
for a sufficiently wide interval of $\lambda.$ 
Note that the ground states of \EQ{eq:ssham} are the extreme spin-squeezed states studied in \REF{Sorensen2001Entanglement}. 
Then, the points of the curve $F_J(X)$ are obtained as $X=\tfrac 1 {J} \aver{L_z}_{\phi_\lambda}$ and $F_J(X)=\tfrac 1 {J}\aver{L_x^2}_{\phi_\lambda}.$  
Note that the method takes into account that the state minimizing $\va{J_x}$ for a given $\exs{L_z}$ has $\exs{L_x}=0,$ which is a property numerically observed to be true for integer $J$ \cite{Sorensen2001Entanglement}.
The algorithm can be extended to half-integer $J$'s by adding a Lagrange multiplier term $\lambda_2 L_x$ that constraints $\exs{L_x}$ to some value, the details can be found in \ref{sec:app_halfinteger}.
In practice, $F_J(X)$ is computed typically for an integer $J$ only, which makes it possible to detect $(k+1)$-particle entanglement for any $k$ for an integer $j$ and for an even $k$ for a half-integer $j.$
In the latter case, it is not a large restriction to consider only even $k,$ since the entanglement depth in 
cold atom experiments can be quite large \cite{Gross2010Nonlinear,Lucke2014Detecting,Hosten2016Measurement}. 

\begin{figure}
\hskip2.5cm\includegraphics[width=0.75\linewidth]{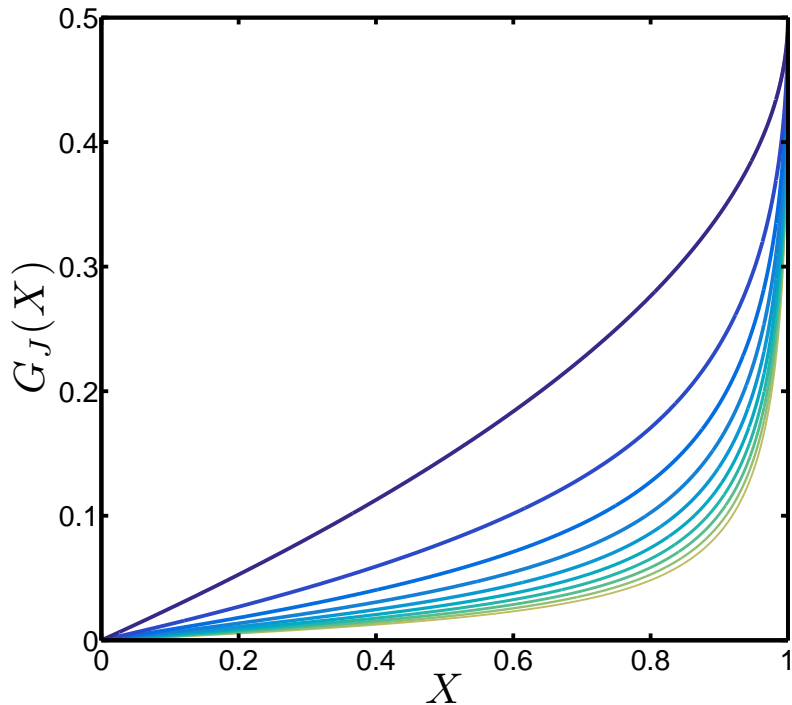}
\caption{The function $G_J(X)$ defined in  \EQ{GJ} for (left to right) $J=1,3,5,\dots,19$.}\label{fig:fsqrtplot}
\end{figure}

In a similar fashion, we can also obtain the curve for $G_J(X)$ defined in \EQ{GJ}.
The points of the curve are given as $X=\tfrac 1 {J^2} \aver{L_z}^2_{\phi_\lambda}$ and $G_J(X)=\tfrac 1 {J}\aver{L_x^2}_{\phi_\lambda}.$ In \FIG{fig:fsqrtplot}, we drew
$G_J(X)$ for various values of $J.$ 
Based on these, the boundary 
for $k$-producible states in the $(\aver{J_y^2+J_z^2},(\Delta J_x)^2)$-plane is given by 
\begin{eqnarray}
\aver{J_y^2+J_z^2}_\lambda&=&\frac{N (N-k)j^2}{k^2j^2} \aver{L_z}^2_{\phi_\lambda}+Nj(kj+1),\nonumber\\
(\Delta J_x)_\lambda^2&= &\frac N {k} (\Delta L_x)_{\phi_\lambda}^2.
\end{eqnarray}
In the numerical calculations, $L_l$ are Hermitian matrices of size $(2kj+1)\times(2kj+1).$
Hence, it is possible to draw the boundaries for various levels of multipartite entanglement for $kj$ reaching up to the thousands, and for an arbitrarily large $N.$

\begin{figure}
\hskip2.5cm\includegraphics[width=0.75\linewidth]{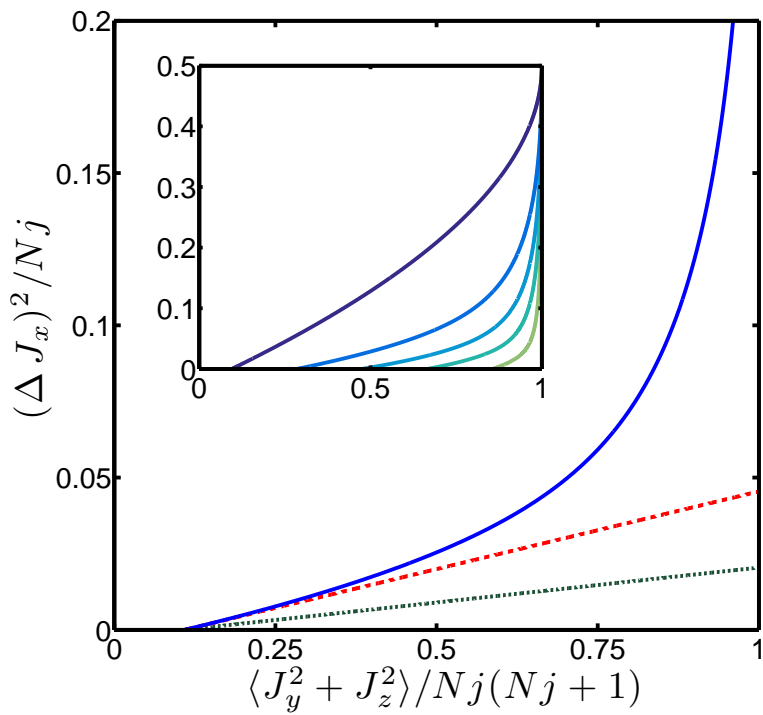}
\caption{$20$-producibility criteria for $N=200$ qubits. (solid) The boundary given by \EQ{eq:oursormolcritintro}. 
(dashed) Criterion \eqref{eq:ourparameter}, i.e., the tangent to the curve given by \EQ{eq:oursormolcritintro}. (dotted) Criterion (\ref{eq:duancon}) given in \REF{Duan2011Entanglement}. (inset) Curves for $k$-producibility for $N=20$ spin-$1$ particles, for  (left to right) $k=1,5,9,13,17$.}\label{fig:kprod_tang}
\end{figure}

We mention that for $J=1$ we have $G_1(X) =\frac 1 2 (1-\sqrt{1-X})$, i.e., the function on the right-hand side of the criteria can be obtained analytically. 
Substituting $F_1(X)=G_1(X^2)$ into \EQ{eq:oursormolcritintro}, we can obtain an analytic $2$-producibility condition for qubits and an analytic separability condition for qutrits.
In \FIG{fig:kprod_tang}, we plotted the curves 
for $k$-producible states for some examples with spin-$\frac 1 2$ and spin-$1$ particles.
For higher $J$, the function $G_{J}(X)$ is not known analytically.
Based on uncertainty relations of angular momentum operators, a lower bound on $G_{J}(X)$ for any $J$ can be obtained as 
\begin{equation}\label{eq:smlowerbound}
\tilde G_{J}(X)= \frac 1 2 \left[ (J+1)-J X - \sqrt{(J+1-J X)^2- X} \right],
\end{equation}
which is not tight for small $J$ and small $X$, but becomes 
tight for large $J$ and $X$ close to $1$
\cite{Sorensen2001Entanglement}.

\section{Nonlinear criterion}

In this section, we present our first main result.

\vskip 0.4cm
\noindent {\bf Observation 1.} The inequality in \EQ{eq:oursormolcritintro} holds for all $k$-producible states of $N$ spin-$j$ particles.
Thus, every state of $N$ spin-$j$ particles that violates \EQ{eq:oursormolcritintro} must be $(k+1)$-entangled. 
The condition \eqref{eq:oursormolcritintro}  can be used if $\aver{J_y^2+J_z^2}\geq Nj(kj+1)$, while
otherwise there is a  $k$-producible quantum state for which $(\Delta J_x)^2=0.$ 

\vskip 0.4cm
\noindent  {\bf Proof.} The key argument of the proof is that for pure $k$-producible states 
of $N$ spin-$j$ particles
\begin{equation}\label{eq:condextrssfoll}
\frac{\sqrt{\aver{J_y}^2+\aver{J_z}^2}}{Nj} \geq \sqrt{\frac{\aver{J_y^2 + J_z^2 }-Nj \left(kj + 1\right)}{N(N-k)j^2}} 
\end{equation}
holds, which is proven in \ref{sec:app_proof_of_formula}. 
Then, based on \EQ{eq:condextrssfoll} and on the monotonicity of $F_J(X)$ in $X,$ we have
for pure $k$-producible states 
\be
\label{RHSLHS1}
F_{J}( {\rm LHS}) \geq F_{ J}( {\rm RHS}).
\ee 
Here, we used the notation ${\rm LHS}$ and ${\rm RHS}$ for the left-hand side and right-hand side of the relation
\EQ{eq:condextrssfoll}, respectively.
On the other hand, the S\o rensen--M\o lmer criterion \eqref{sormolcrit} can be rewritten as
\be
\label{RHSLHS2}
(\Delta J_x)^2 \geq NjF_{J}( {\rm LHS}).
\ee 
From \EQ{RHSLHS1} and \EQ{RHSLHS2}  follows that \EQ{eq:oursormolcritintro} holds for pure $k$-producible states. 

Next, we will consider mixed states.
In the formula \eqref{eq:oursormolcritintro} the argument of $G$ is linear in the state.
Then, our criterion \eqref{eq:oursormolcritintro} can be extended to mixed $k$-producible states via a convex hull of $G_{J}(X)$. However we can observe numerically that $G_{J}(X)$ is convex already by itself. The tightness of \EQ{eq:oursormolcritintro} is discussed in \ref{sec:app_tightness}, while the convexity of $G_{J}(X)$ is considered in detail in \ref{sec:app_profFJGJ}.\qed

\vskip 0.4cm

The criterion (\ref{eq:oursormolcritintro}) is especially suited to detect states for which $\aver{J_y^2 + J_z^2}$ is large and $(\Delta J_x)^2$ is small. A paradigmatic example for such a state is the unpolarized Dicke state in the $x$-basis 
\be 
\label{eq:Dickex} 
\rho_{\rm Dicke}=|J=Nj,m_x=0\rangle\langle J=Nj,m_x=0|,
\ee 
that satisfies $(\Delta J_x)^2=0$ and $\aver{J_y^2 + J_z^2}=Nj(Nj+1)$ and is detected as $N$-entangled. In fact, substituting these quantities in the criterion \EQ{eq:oursormolcritintro} the left-hand side is zero, while the right-hand side is positive for $k=N-1.$ Note also that the Dicke states violate maximally even the relation \eqref{eq:condextrssfoll} for pure $k$-producible states \footnote{We stress that \EQ{eq:condextrssfoll} is not an entanglement criterion, since it does not hold for mixed  states.}. 

\section{Linear analytic criteria}

In this section, we will derive the spin-squeezing parameters \EQ{eq:ourparameter} and \EQ{eq:sormolparameter}. 
Complementary to the approximation \EQ{eq:smlowerbound}, our approach is based on a lower bound on $G_{J}(X)$ that is tighy for
$X\approx 0$ and improves $\tilde G_{J}$ at small $X$ by a factor~of~$2$. 
For our derivation, we will compute the first terms of the Taylor expansion of $G_J(X)$ around $X=0.$ 
Using the convexity of $G_J(X),$ we will obtain the bound $G_J(X) \geq (G_J(0) + X  G^\prime_J(0)),$ with $G_J(0)=0.$ In other words, we will compute the tangent to the $k$-producibility boundaries, near their intersection point with the horizontal axis.

In what follows, we present the details of the derivation. The expansion of  $G_J(X)$ can be obtained by employing the perturbation series for $H_\lambda$
in powers of the parameter $\lambda \ll 1$, taking advantage of the fact that $X=0$ corresponds to $\lambda=0$. 
The ground state of $H_\lambda$ is then given by
 $|\phi_\lambda\rangle= |\phi^{(0)}\rangle+\lambda |\phi^{(1)}\rangle+O(\lambda^2)$
 \footnote{$O(x)$ is the usual Landau  symbol used to describe the asymptotic behavior of
a quantity for small $x$}, where $|\phi^{(0)}\rangle$ is the ground state of the unperturbed Hamiltonian $H^{(0)}=L_x^2$, i.e., the eigenstate of $L_x$ with eigenvalue zero. 
As in usual perturbation theory, the first order term is obtained by imposing $\langle\phi^{(0)} |\phi^{(1)}\rangle=0$ and results in $|\phi^{(1)}\rangle= \sum_{m\neq 0} c_{m} | E^{(0)}_m \rangle$, where $c_m=-\langle E^{(0)}_m| H^{(1)} | E^{(0)}_0\rangle / (E^{(0)}_m-E^{(0)}_0)$ and $E^{(0)}_m$ are the energy levels of the unperturbed Hamiltonian. In our case, we obtain 
$|\phi^{(1)}\rangle = \sum_{m\neq 0} c_{m} | m \rangle_x$ with $c_{m}= -\langle m|_x L_z | 0\rangle_x / m^2,$ where $| m\rangle_x$ are the eigenstates of $L_x$ with eigenvalue $m.$
The expansion of the ground state explicitly is as follows
\begin{equation}
\label{eq:approxground}
|\phi_\lambda\rangle = | 0 \rangle_x -i\lambda \tfrac{\sqrt{J(J+1)}}{2}\left(| 1 \rangle_x - | -1 \rangle_x\right) + 
O(\lambda^2).
\end{equation}
Based on \EQ{eq:approxground}, we obtain for the dependence of $X$ and $G_J(X)$ on $\lambda,$ respectively, 
$X(\lambda)=\tfrac 1 {J^2} \aver{L_z}^2_{\phi_\lambda}\approx \lambda^2 (J+1)^2$ and $G_J(X(\lambda))=\tfrac 1 J \aver{L_x^2}_{\phi_\lambda}\approx \tfrac 1 2 \lambda^2 (J+1).$ Hence, we arrive at 
\begin{eqnarray}
\label{eq:taylorfj}
G_J(X) &\geq& \frac{X}{2(J+1)} , 
\end{eqnarray}
by employing the chain rule for $\tfrac{{\rm d} G_J(X(\lambda))}{{\rm d} X}$
 near $X=\lambda=0.$ 
Based on this, we can derive an analytic criterion that becomes  tight close to the point $(\Delta J_x)^2~=~0$. 
Note that we could also use  $\tilde G_J(X)$ defined in \EQ{eq:smlowerbound} instead of  $G_J(X)$  for constructing our linear entanglement condition. However, taking   the derivative of $\tilde G_J(X)$   one obtains $G_J(X) \geq \tilde G_J(X) \geq X \tilde G^\prime_J(0)= \tfrac{X}{4(J+1)}$, which underestimates \EQ{eq:taylorfj} by a factor of 2.
Note  that we computed the leading terms for the Taylor expansion of $G_J(X)$ analytically, while the function itself is known only numerically. 

\vskip 0.4cm
\noindent  {\bf Observation 2.} The criteria in \EQ{eq:ourparameter} and \EQ{eq:sormolparameter}
hold for all $k$-producible states of $N$ spin-$j$ particles such that $J$ given in \EQ{eq:Jkj} is an integer number.
Every state of $N$ spin-$j$ particles that violates one of the criteria must be $(k+1)$-entangled, i.e.,
has an entanglement depth at least $k+1.$
\vskip 0.4cm

\noindent  {\bf Proof.} From \EQ{eq:taylorfj} 
we can bound the criterion \eqref{eq:oursormolcritintro} from below  with \EQ{eq:ourparameter} by substituting $X=\left[\aver{J_y^2+J_z^2} - Nj(kj+1)\right]/[N (N-k)j^2]$. Analogously, by 
rewriting \EQ{sormolcrit} in terms of $G_J$ and using the bound (\ref{eq:taylorfj}) with $X~=~\aver{J_z}^2 / N^2j^2$ we obtain \EQ{eq:sormolparameter}.~\qed
\vskip 0.4cm

In \FIG{fig:kprod_tang}, we plot the criterion~(\ref{eq:ourparameter}) as the tangent to the boundary of $20$-producibility for $N=200$ particles with spin $j=\frac 1 2$ in the $(\aver{J_y^2+J_z^2},(\Delta J_x)^2)$-plane.

\section{\bf Comparison with similar criteria}

Next, we compare our criteria with other similar entanglement conditions. First let us consider the S\o rensen-M\o lmer criterion \eqref{sormolcrit}. 

\vskip 0.4cm
\noindent  {\bf Observation 3.}  Whenever the condition 
\begin{equation}\label{eq:ourstonger}
\frac{(\Delta J_y)^2 + (\Delta J_z)^2}{Nj}  > kj \left(1- \frac{\aver{J_y}^2 + \aver{J_z}^2}{N^2j^2} \right) +1 
\end{equation}
holds then our criterion~(\ref{eq:oursormolcritintro}) is strictly stronger than
the S\o rensen--M\o lmer criterion (\ref{sormolcrit}). 

\vskip 0.4cm
\noindent  {\bf Proof.} (a) Since $F_{J}(X)$ is a monotonously increasing function of $X,$  the inequality $F_{J}(X) \geq F_{J}(Y)$ holds if and only if $X \geq Y.$ Hence, for comparing \EQ{sormolcrit} and \EQ{eq:oursormolcritintro} it suffices to compare the arguments of the function $F$ in the two conditions.
It is then straightforward to prove that
 \EQ{eq:oursormolcritintro} implies
\EQ{sormolcrit} whenever \EQ{eq:ourstonger} holds.
Then, let us now present a family of states that are detected by \EQ{eq:oursormolcritintro}, but not detected by \EQ{sormolcrit}.
We consider states of the form \be\label{eq:noisyDicke}\rho_{{\rm Dicke}, p}~=~(1-p)\rho_{\rm Dicke}+p\frac{\id}{(2j+1)^N},\ee where the unpolarized Dicke state is given in \EQ{eq:Dickex}.  From the linear criterion (\ref{eq:ourparameter}) we obtain that if  $p<\frac{3(N-k)j}{2j(j+1)(kj+1)(N-k)-2(j+1)+3(Nj+1)}$ then the state $\rho_{{\rm Dicke}, p}$ is detected by \EQ{eq:oursormolcritintro}. On the other hand, $\rho_{{\rm Dicke}, p}$ is not detected  by the S\o rensen--M\o lmer criterion~(\ref{sormolcrit}), since $\langle J_l\rangle=0$ for $l=x,y,z$ for this state for all $p.$ \qed
\vskip 0.4cm

From Observation 3, 
we can immediately see that our criterion \EQ{eq:oursormolcritintro} is much stronger than the original spin-squeezing criterion \eqref{sormolcrit}
for states close to Dicke states \eqref{eq:Dickex}  since for such states $(\Delta J_y)^2 + (\Delta J_z)^2 \gg Nkj^2.$ Here, we  assumed that 
 $k$ is much smaller than $N,$ which is consistent with experiments, where criterion \eqref{sormolcrit}  always detects an entanglement depth much smaller than $N$  due to noise \cite{Gross2010Nonlinear,Hosten2016Measurement}. 

Let us now study numerically how our criterion works for a relevant class of states.
We consider spin-squeezed states of spin-$\tfrac 1 2$ particles obtained from ground states of the Hamiltonian
\be 
H_\mu=J_x^2-\mu J_z,
\label{eq:Hmu}
\ee
for simplicity assuming an even particle number.
The Dicke state \eqref{eq:Dickex} corresponds to $\mu=0,$ while
the usual spin-squeezed states with a large spin polarization correspond to a large $\mu.$
For such states without noise, our criterion \EQ{eq:oursormolcritintro} is not stronger than \EQ{sormolcrit}.

\begin{figure}
\hskip2.5cm
\includegraphics[width=0.75\linewidth]{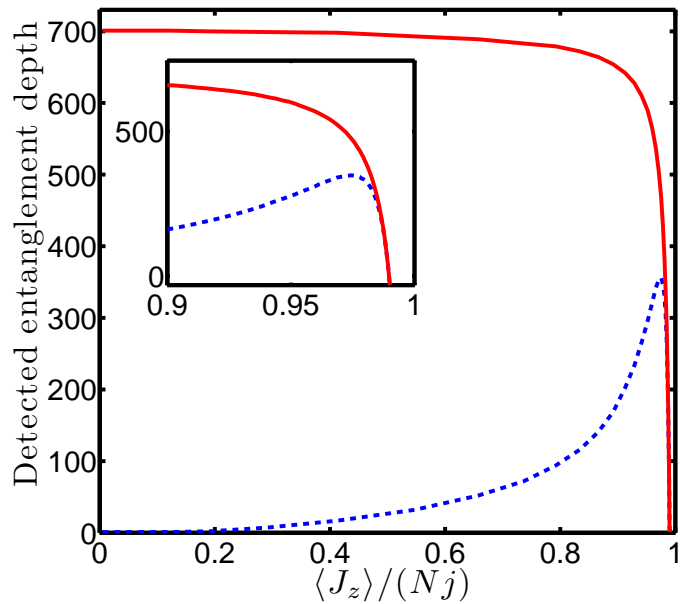}
\caption{Multiparticle entanglement for spin-squeezed states of $N=1000$ spin-$\tfrac 1 2$ particles, after 10 particles decohered into the completely mixed state.
(solid) Entanglement depth detected by our criterion \eqref{eq:oursormolcritintro}  and (dashed) the S\o rensen-M\o lmer criterion \eqref{sormolcrit}.  }\label{fig:2criteria}
\end{figure}

Simple calculations show that if some small noise is present in the system then \EQ{eq:oursormolcritintro}  detect an entanglement depth higher  than the original criterion \eqref{sormolcrit}. 
First we consider spin-squeezed states for $N=1000$ spin-$\tfrac 1 2$ particles, such that $10$ particles are decohered into the fully mixed state. Such a noise is typical in cold atom experiments \cite{Echaniz2005Conditions}. The results can be seen in \FIG{fig:2criteria}. 
Our criterion \eqref{eq:oursormolcritintro} and  the S\o rensen-M\o lmer criterion \eqref{sormolcrit} detect the same entanglement depth for almost completely polarized spin-squeezed states. On the other hand, as the squeezing increases, our criterion detects a monotonically increasing entanglement depth, while the other criterion detects smaller and smaller multipartite entanglement. 
While we considered a noise affecting a few particles, still the detected entanglement depth is much smaller than $N.$
Other types of noise, such as particle loss, small added white noise, or noise effects modelled considering the thermal states of \EQ{eq:Hmu} lead to a similar situation.

Next, we compare our criteria with another  important  condition that is  designed to detect the depth of entanglement near unpolarized Dicke states \eqref{eq:Dickex}. 
It is a linear criterion derived by Duan in 
\REF{Duan2011Entanglement}, stating that
\begin{equation}\label{eq:duancon}
N(k+2)(\Delta J_x)^2 \geq \aver{J_y^2 + J_z^2}-\frac N 4 \left(k + 2\right)
\end{equation}
holds for all $k$-producible states of $N$ spin-$\frac 1 2$ particles. Any state that violates \EQ{eq:duancon} is detected as $(k+1)$-entangled. In this case, we can compare it with the linear criterion (\ref{eq:ourparameter}), specialized to qubit-systems, i.e., for $j=\tfrac 1 2$  
\begin{equation}\label{eq:ourtangqubits}
\frac{(N-k)} 2 \left(k+2 \right) (\Delta J_x)^2 \geq \aver{J_y^2 + J_z^2} - \frac N 4 \left(k+2 \right) .
\end{equation}
It is easy to see that a violation of \EQ{eq:duancon} implies a violation of \EQ{eq:ourtangqubits}. Thus, our condition detects more states, which can be seen in \FIG{fig:kprod_tang}.

Finally, we note that \EQ{eq:oursormolcritintro} with $j=1/2$ is similar to the criterion for spin-$1/2$ particles used in the experiment described in \REF{Lucke2014Detecting}. A key difference is that 
in \EQ{eq:oursormolcritintro}, in the denominator of the fraction, the term $N (N-k)j^2=N(N-k)/4$ appears, while in the formula of \REF{Lucke2014Detecting} there is the term $N^2/4.$ The difference between the two criteria is the largest when we examine highly entangled Dicke states or spin-squeezed states, and in the argument of $F_J(X)$ we have a value close to $X=1.$ In the vicinity of this point, the derivative of  $F_J(X)$ is very large, hence some improvement
in the argument of $F_J(X)$ makes the bound on the right-hand side of \EQ{eq:oursormolcritintro} significantly higher. 
As a consequence, the criterion \eqref{eq:oursormolcritintro} 
can be used to detect  
the noisy Dicke states of many particles even in $k\sim N$ case, while the criterion of \REF{Lucke2014Detecting} 
can be used only when $k\ll N,$ and it does not detect the Dicke state as $N$-entangled.

\section{Extension to fluctuating number of particles}

For macroscopic ensembles of particles, e.g., for $N\sim 10^6$, the total particle number is not under perfect control. In this section, we will generalize our entanglement criteria  to such a situation.
The quantum state of a large particle ensemble with a fluctuating particle number is given as \be\label{eq:rhofluct}\rho=\sum_{N} Q_{N} \rho_{N},\ee where $\rho_{N}$ are the density matrices corresponding to a subspace with a particle number $N$ and $Q_{N}$ are probabilities.
We also have to consider collective spin operators defined as $J_l = \sum_{N} J_{l, N}$ for $l=x,y,z,$ where 
 $J_{l, N},$ act on the subspace with $N$ particles. In principle, one could evaluate an entanglement condition, e.g., 
\EQ{sormolcrit} for one of the fixed-$N$ subspaces.  If $\rho_{N}$ has an entanglement depth $k$ for some $N,$ then the state $\rho$ has also at least an entanglement depth $k.$ 
However, in practice, we would not have sufficient statistics to evaluate our entanglement criteria for some fixed $N.$ This issue has been studied by Hyllus \emph{et al.} \cite{Hyllus2012Entanglement}, who generalized the  definition of entanglement depth to the case of a fluctuating number of particles.  
They also showed how spin-squeezing criteria can be used in this case such that all the collected statistics is used, not only data for a given particle number $N.$
For instance, \EQ{sormolcrit} can be transformed to  \cite{Hyllus2012Entanglement}
\begin{equation}\label{sormolcritfluctN}
(\Delta J_x)^2 \geq \aver{N}j F_{ J}\left( \frac{\aver{J_z}}{\aver{N}j} \right) .
\end{equation}
Here, \EQ{sormolcritfluctN} could be obtained from  \EQ{sormolcrit} simply
with the substitution $N \rightarrow \aver{N}.$

Using methods similar to the ones in \REF{Hyllus2012Entanglement}, 
we will now obtain the criterion \eqref{sormolcrit} for fluctuating particle numbers.

\vskip 0.4cm
\noindent  {\bf Observation 4.}  All $k$-producible states with a fluctuating particle number
must satisfy the
following inequality
\begin{equation}\label{eq:oursormolcritN}
(\Delta J_x)^2 \geq  \aver{N}j  G_{J}\left(\frac{\aver{W}}{\aver{N}j}\right) ,
\end{equation}
where we define the operator
\begin{equation}\label{eq:woper}
W=\sum_N (Nj-kj)^{-1}\left[ J_{y,N}^2+J_{z,N}^2 -Nj(kj+1)\id_N \right] ,
\end{equation} 
$\aver{W}\geq 0$ is required, and ${J}$ is the total spin of a $k$-particle group given in \EQ{eq:Jkj}. 
\vskip 0.4cm
\noindent  {\bf Proof.}  We have to start from a state of the form \eqref{eq:rhofluct}. Due to the concavity of the variance,
the variance of the mixed state can be bounded from below with the variances of $\varrho_N$ as $(\Delta J_k)^2 \geq \sum_{N} Q_{N} (\Delta J_{k, N})^2.$
Moreover, since $G_J(X)$ is convex in $X,$ it has to satisfy Jensen's inequality. Thus,
\be\tfrac 1 {\sum_{N} Q_{N} N} \sum_{N} Q_{N} N G_{ J}\left( X_N \right) \geq  G_{ J}\left(\tfrac{\sum_{N} Q_{N} N X_N}{\sum_{N} Q_{N} N}\right)\ee with $X_N=\tfrac{\aver{W}_{\rho_N}}{N}$~and~$\aver{N}~=~\sum_{N} Q_{N} N$ must hold.
Based on these, the statement of the Observation follows. \qed
\vskip 0.4cm

Note that the operator $W$ defined in \EQ{eq:woper} is simply a sum of $J_{y,N}^2+J_{z,N}^2$ over all fixed-$N$ subspaces, normalized with $(Nj-kj)$. Thus, to apply our condition in experiments with fluctuating number of particles, one needs to measure the spin operators and the particle number jointly at each shot, and then average over an ensemble. 

Finally, let us consider how to apply the ideas above for the spin-squeezing parameters
defined in this paper. The parameter \eqref{eq:sormolparameter} can be extended to fluctuating particle numbers simply by replacing 
$N$ with $\exs{N}.$
Similarly, for the parameter \eqref{eq:ourparameter}, we have to replace
$\frac{\aver{J_y^2+J_z^2}-Nj(kj+1)}{(N-k)j}$ by $\exs{W}.$

\section{Conclusions}
We derived a set of criteria to determine 
the depth of entanglement of spin-squeezed states and unpolarized Dicke states, extending and completing the results of Refs.~\cite{Sorensen2001Entanglement,Lucke2014Detecting}. These generalized spin-squeezing conditions are valid even for an ensemble of  spin-$j$ particles with $j>\frac 1 2,$
which is very useful, since most experiments are carried out with particles with a higher spin,  e.g., with spin-$1$ $^{87}{\rm Rb}$ atoms.
Since theory is mostly available for the spin-$\frac 1 2$ case, pseudo spin-$\frac 1 2$ particles are created artificially such that only two of the levels are populated.
While the spin-squeezing approach to entanglement detection is already widely used in such systems \cite{Hald1999Spin,Fernholz2008Spin,Riedel2010Atom-chip-based,Orzel2386,Esteve2008Squeezing,Gross2010Nonlinear,Hamley2012Spin-nematic,Behbood2014Generation,BohnetJ2014,Lucke2014Detecting,Cox2016},
our criteria make it possible to study spin-squeezing in fundamentally new experiments. A clear advantage of using the physical spin is that it is typically much easier to manipulate than the  pseudo spin-$\frac 1 2$  particles \cite{Behbood2014Generation}. In future, it would be interesting to clarify the relation between generalized spin squeezing and metrological usefulness \cite{Hyllus2012Fisher,Toth2012Multipartite,Zhang2014Quantum,Toth2014Quantum,Apellaniz2015Verifying,Apellaniz2015Optimal}, and also compare our results with the complete set of spin-squeezing criteria of \REF{Vitagliano2011Spin}, which contain one additional collective observable, related to single-spin average squeezing.
\ack
We thank G. Colangelo, L. M. Duan, O. G\"uhne, P. Hyllus, M. W. Mitchell, and J. Peise for discussions. This work was supported by the EU (ERC Starting Grant 258647/GEDENTQOPT, CHIST-ERA QUASAR, COST Action CA15220)
the Spanish MINECO
(Project No. FIS2012-36673-C03-03 and No. FIS2015-67161-P),
the Basque Government (Project No. IT4720-10),
the National Research Fund of Hungary OTKA (Contract No. K83858),
the DFG (Forschungsstipendium KL 2726/2-1), and the FQXi (Grant No. FQXi-RFP-1608).
We also acknowledge support from the Centre QUEST, the DFG through RTG 1729 and
CRC 1227 (DQ-mat), project A02, and the EMRP.

\appendix

\section{Computing $F_J(X)$ and $G_J(X)$ for half-integer spin}
\label{sec:app_halfinteger}

For half-integer spins, we have to calculate $F_J(X)$ numerically as follows. We consider the Hamiltonian \cite{Sorensen2001Entanglement}
\begin{equation}\label{eq:ssham2lambdas}
H_{\lambda,\lambda_2}=L_x^2-\lambda L_z -\lambda_2 L_x,
\end{equation}
and denote its ground state by $\phi_{\lambda,\lambda_2}.$ Then, $F_J(X)$ can be obtained as
\begin{equation}
F_J(X)=\min_{\lambda,\lambda_2:\frac{1}{J}\exs{L_z}=X} (\Delta J_x)^2_{\psi_{\lambda,\lambda_2}},
\end{equation}
which is a two-parameter optimization with the constraint $\frac{1}{J}\exs{L_z}=X.$

\section{Details of the proof of Observation 1}

\subsection{Proof. of \EQ{eq:condextrssfoll}}
\label{sec:app_proof_of_formula} 

To prove \EQ{eq:condextrssfoll}, let us consider the expression $(\Delta J_y)^2+(\Delta J_z)^2$ on pure $k$-producible states \eqref{eq:kprod}.
Due to the additivity of the variance for tensor products
\begin{eqnarray}\label{eq:saturatable}
&&(\Delta J_y)^2+(\Delta J_z)^2 = \sum_{l} \left[(\Delta j_y^{(l)})^2+(\Delta j_z^{(l)})^2\right] \nonumber\\
&&\;\;\;\;\;\;\;\;\;\;\;\;\;\;\leq \sum_{l} \left[ k_l j \left(k_l j + 1\right) - \aver{(j_x^{(l)})^2}- \aver{j_y^{(l)}}^2-\aver{j_z^{(l)}}^2\right] 
\end{eqnarray}
holds, where the superscript $(l)$ indicates the $l$th group, that is composed of $k_l$ particles. 
The inequality \eqref{eq:saturatable} is saturated by all quantum states for which
$\aver{(j_x^{(l)})^2+(j_y^{(l)})^2+(j_z^{(l)})^2}$ is maximal, i.e., equal to $k_l j \left(k_l j + 1\right)$, for all $l$.

For simplifying our expression, we neglect the non-negative  quantity  
\be\label{eq:mathcalX}
\mathcal X:=\sum_l \aver{(j_x^{(l)})^2},
\ee 
and after some rearrangement of the terms in \EQ{eq:saturatable} we arrive at
\begin{eqnarray} \label{eq:saturatable2} 
&&\aver{J_y^2 + J_z^2} \leq \aver{J_y}^2 + \aver{J_z}^2 \nonumber\\
&&\;\;\;\;\;\;\;\;\;\;\;\;\;\;+\sum_{l} k_l j  \left[ \left(k_l j + 1\right)-k_l j \frac{\left(\aver{j_y^{(l)}}^2+\aver{j_z^{(l)}}^2\right)}{k_l^2j^2}\right] .
\end{eqnarray}
From \EQ{eq:saturatable2}, we can obtain a simpler bound as 
\begin{eqnarray}\label{eq:supps3}
&\aver{J_y^2 + J_z^2}\leq \aver{J_y}^2 + \aver{J_z}^2+ Nj \nonumber\\
&\;\;\;\;\;\;\;\;\;\;\;\;\;\;+ \sum_{l} k_l j  \left[ k j \left( 1- \frac{\aver{j_y^{(l)}}^2+\aver{j_z^{(l)}}^2}{k_l^2j^2}\right) \right] ,
\end{eqnarray}
due to the fact that $k_l\le k, \sum_l k_l=N,$ and that the expression inside the round brackets in \EQ{eq:supps3} is positive. Furthermore, using Jensen's inequality in the form
\begin{equation}\label{jens1}
- \sum_{l} k_l f_l^2 \leq - \frac 1 N \left(\sum_{l} k_l f_l \right)^2  , \ \sum_{l} k_l = N , 
\end{equation}
with $f_l=\frac{\aver{j_{m}^{(l)}}}{k_{l}}$ for $m=x,y,z$ we obtain
\begin{equation}\label{eq:ineqJyJzvarJx}
\aver{J_y^2 + J_z^2}-Nj(kj+1)
\leq  \left(1-\frac k N \right) (\aver{J_y}^2 + \aver{J_z}^2) .
\end{equation}
Hence, we proved \EQ{eq:condextrssfoll}.

\subsection{Tightness of \EQ{eq:condextrssfoll} and \eqref{eq:oursormolcritintro}}
\label{sec:app_tightness}

We will now examine, how the relation \eqref{eq:condextrssfoll} would look for pure $k$-producible states \eqref{eq:kprod} without neglecting $\mathcal X$ defined in \EQ{eq:mathcalX}. Simply, $\aver{J_y^2 + J_z^2}$ would be substituted by $\aver{J_y^2 + J_z^2}+\mathcal X.$
With a derivation similar to the one in \ref{sec:app_proof_of_formula}, it can be shown that such a condition would be saturated by all quantum states of the form $\ket{\psi} \otimes \ket{\psi}\otimes ... \otimes\ket{\psi},$ if $\ket{\psi}$ are $k$-qubit states and $\aver{j_x^2+j_y^2+j_z^2}_{\psi}$ is maximal, i.e., it is $k j \left(k j + 1\right).$ (Here we assumed that $\mathcal X$ is defined such that all particle groups contain $k$ particles, i.e, $k_l=k$ for all $l.$)

Let us now see how large $\mathcal X$ is for relevant states.
For the state fully polarized in the $z$-direction, we have \be\mathcal X=\sum_l \va{j_x^{(l)}}= Nj^2/2,\ee where we used the fact that $\exs{j^{(l)}}=0$ for such a state. Let us consider now the ground states of the Hamiltonian \eqref{eq:Hmu} for a given parameter $\mu.$ Such states include usual spin-squeezed states, as well as Dicke states \eqref{eq:Dickex}. For any $\mu,$ \be\label{eq:upperbound}\mathcal X < Nj^2/2\ee holds, since for such states the variance of the $x$-components of the collective angular momentum is squeezed below that of the completely polarized state for any particle group. Note that the upper bound in \EQ{eq:upperbound} does not grow with $k.$

Let us now consider the other relevant quantity, $\exs{J_y^2+J_z^2}.$ For the state fully polarized in the $x$-direction, we have $\exs{J_y^2+J_z^2}=Nj(Nj+1/2).$ For the Dicke state \eqref{eq:Dickex}, $\exs{J_y^2+J_z^2}=Nj(Nj+1).$ For ground states of \EQ{eq:Hmu} for $\mu>0,$ $\exs{J_y^2+J_z^2}$ is in between these two values. This can be seen noticing that
$\exs{J_x^2+J_y^2+J_z^2}=Nj(Nj+1)$ for these states.

Based on the previous discusion, it is clear that $\mathcal X\ll \exs{J_y^2+J_z^2}$ holds for large $N.$ Hence, in practical cases the relation \eqref{eq:oursormolcritintro} provides a tight bound on $\va{J_x}$ based on $\exs{J_y^2+J_z^2}$ in the large $N$ limit.

\subsection{Properties of $F_{J}$ and $G_{J}$}
\label{sec:app_profFJGJ}

The functions $F_{J}(X)$ can be obtained from the optimal states $\rho$ for the problem defined in \EQ{fsormolthj},
i.e., the states that minimize $(\Delta L_x)^2$ for a given $\aver{L_z}$. As discussed in \SEC{sec:numcomp}, 
for an integer spin $J$,  such states are the ground states of \EQ{eq:ssham}, where $\lambda$ is a parameter.
They have  $\langle L_x\rangle=0$ \cite{Sorensen2001Entanglement}. Thus, $F_J(X)$ yields the minimal $\aver{J_x^2}$ for a given value of $\aver{J_z}$. Since the set of physical states is convex, the set of points in the $(\aver{J_z},\aver{J_x^2})-$space corresponding to physical states is also convex. Hence, $F_J(X)$ is also a convex function and in particular its derivative $\lambda(X)$ is monotonously increasing with $X$. Note that in \REF{Sorensen2001Entanglement} a different proof was presented for this fact. In principle, the derivative $F'_J(X)$ can be computed by numerical derivation of  $F_J(X).$ However, it is much simpler to obtain $F'_J(X)$ for some range of $X$ by plotting $(\tfrac 1 {J} \aver{L_z}_{\phi_\lambda},\lambda)$ for some range of $\lambda$ \cite{Sorensen2001Entanglement}.
In other words, for $X=\tfrac 1 {J} \aver{L_z}_{\phi_\lambda}$ the derivative is $F_J'(X)=\lambda.$

\begin{figure}
\hskip2.5cm\includegraphics[width=0.75\linewidth]{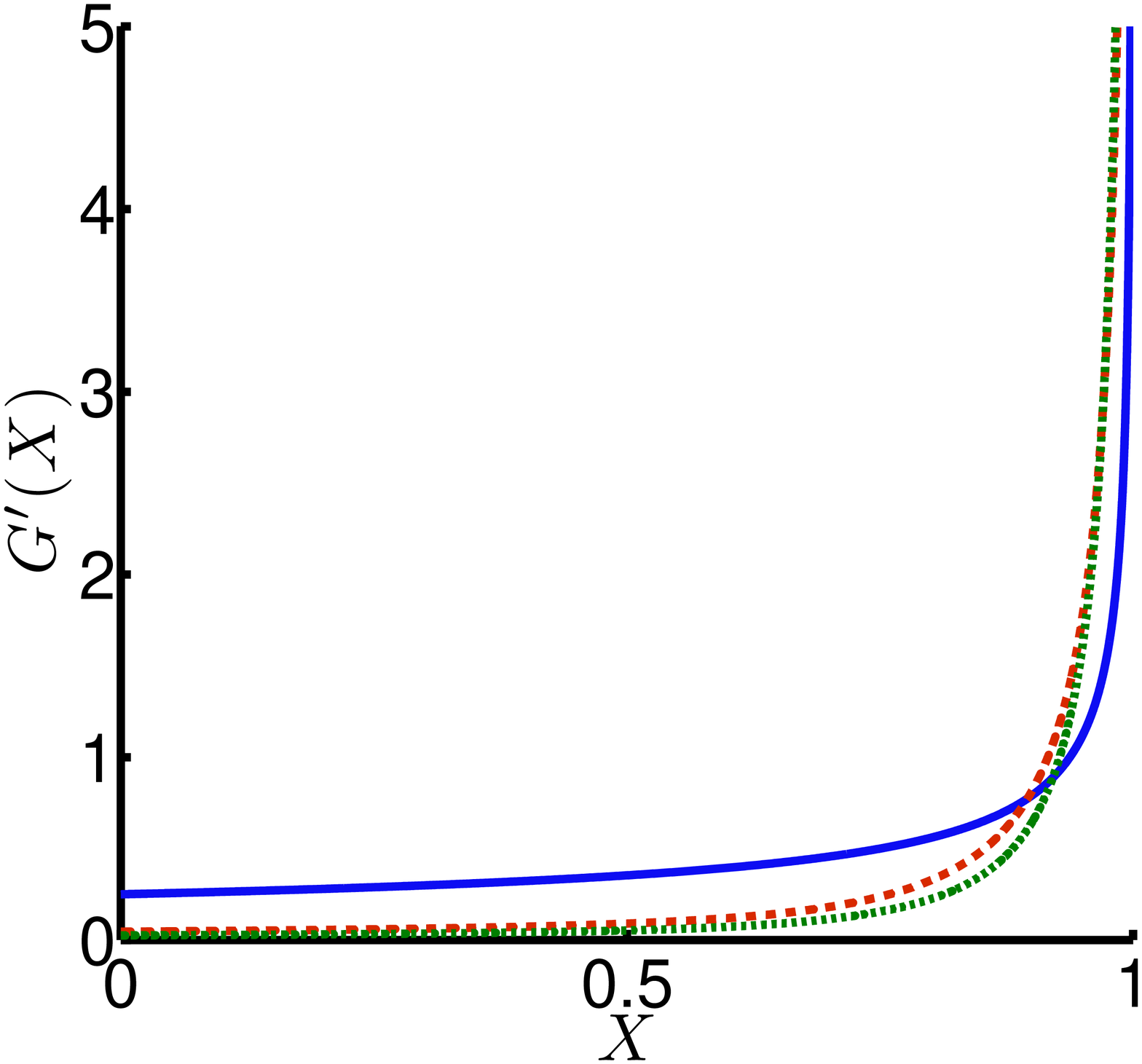}
\caption{The derivative $G^\prime(X)=\frac{J}{2\aver{L_z}_{\phi_\lambda}} \lambda$ as a function of $X=\tfrac 1 {J^2} \aver{L_z}_{\phi_\lambda}^2$ for  (solid) $J=1$, (dashed) $J=10,$  and (dotted) $J=19.$ }\label{fig:lambdaofx}
\end{figure}

To show that also $G_J(X)$ is convex we observe that
$G_J^\prime(X) =  \tfrac{1}{2\sqrt X}F_J'(\sqrt X)$ 
is a monotonously increasing function of $X.$ 
We evaluate numerically the derivative $G_J^\prime(X)$ by plotting $(\frac 1 {J^2} \aver{L_z}_{\phi_\lambda}^2,\frac{J}{2\aver{L_z}_{\phi_\lambda}} \lambda)$ for a wide range of $\lambda$, cf. \FIG{fig:lambdaofx}, and see explicitly its monotonicity.
More generally, one can check whether or not $F_J(X^\frac 1 {\alpha})$ is convex for any exponent $\alpha$. 
It can then be observed numerically (not shown) that $F_J(X^\frac 1 {\alpha})$ is not convex
for any $\alpha>2$.

So far we discussed the case of integer spin. For half-integer spin, the ideas mentioned before cannot be used. Then, the derivative of $G_J$  can be obtained via the numerical derivation of $F_J(X).$ Based on numerics, we can make the same statements about the convexity of $G_J(X)$ and $F_J(X^\frac 1 {\alpha})$ as for the case of an integer spin.

\bibliography{Biblio_dicke_no_url}

\end{document}